\documentclass[a4paper,superscriptaddress,floatfix,twocolumn,showkeys,pre, aps ]{revtex4-1}

\usepackage[utf8]{inputenc}
\usepackage[english]{babel}
\usepackage{amssymb}
\usepackage{amsmath}
\usepackage{graphicx}
\usepackage{float}
\usepackage{subfigure}

\usepackage[normalem]{ulem}
\usepackage{color}

\definecolor{mgray}{rgb}{0.3,0.3,0.3}
\definecolor{sigcol}{RGB}{17,97,165}
\definecolor{shadecolor}{RGB}{211,220,238}
\definecolor{edgecolor}{RGB}{17,97,165}

\newcommand{\km}{\left<k\right>}

\newcommand{\Hl}[1]{\textit{#1}}

\newcommand{\eqdot}{\,.}
\newcommand{\eqcomma}{\,,}

\newcommand{\gvec}[1]{\boldsymbol{#1}}

\newcommand{\stst}[1]{\left. #1 \right|_\mathrm{st}}

\renewcommand{\subsubsection}[1]{}

\begin{document}

\author{Kaj-Kolja Kleineberg}
\email{kkl@ffn.ub.edu}
\affiliation{Departament de F\'isica Fonamental, Universitat de Barcelona, Mart\'i i Franqu\`es 1, 08028 Barcelona, Spain}
\author{Mari\'an Bogu\~{n}\'a}
\affiliation{Departament de F\'isica Fonamental, Universitat de Barcelona, 
Mart\'i i Franqu\`es 1, 08028 Barcelona, Spain}
\date{\today}

\title{Competition between global and local online social networks}

\begin{abstract}
The overwhelming success of online social networks, the key actors in the Web 2.0 cosmos, has reshaped human interactions globally. To help understand the fundamental mechanisms which determine the fate of online social networks at the system level, we describe the digital world as a complex ecosystem of interacting networks.
In this paper, we study the impact of heterogeneity in network fitnesses on the competition between an international network, such as Facebook, and local services. The higher fitness of international networks is induced by their ability to attract users from all over the world, which can then establish social interactions without the limitations of local networks. In other words, inter-country social ties lead to increased fitness of the international network. To study the competition between an international network and local ones, we construct a 1:1000 scale model of the digital world, consisting of the 80 countries with the most Internet users.  
Under certain conditions, this leads to the extinction of local networks; whereas under different conditions, local networks can persist and even dominate completely.
In particular, our model suggests that, with the parameters that best reproduce the empirical overtake of Facebook,
this overtake could have not taken place with a significant probability. 
\end{abstract}


\maketitle

\section{Introduction}

The rapid growth of online social networks (OSNs), such as Twitter or Facebook, led to over two billion accounts being active in 2014~\cite{digital_statshot}, connecting over one quarter of the world's population and $72\%$ of online U.S. adults~\cite{few_report}. Bridging the gap between social sciences, and information and communication technologies, OSNs constitute a crucial building block in the development of innovative approaches to the challenges society faces today. However, technological progress over the last decade has dramatically outpaced our understanding of the new systems and their impact on society.

This lack of understanding of the complex dynamics of the digital world reveals the pressing need for a comprehensive and concise theory to describe and model the online world as a set of interacting networks. In this context, the activity of users has become a resource to be fought over and which drives competition in the digital world.
Digital services only persist if they can attract and maintain users' attention. 
Here, we describe the online world as a complex, digital ecosystem in which interacting networks constitute species in competition for survival. 
Within this ecological context, the extinction of a network corresponds to the complete absence of activity, as an entirely passive network cannot function and, more importantly, will not attract new users. 
In a recent study~\cite{ecology20}, we demonstrated that a moderate number of identical networks can coexist in the digital ecosystem, in contrast to the principle of competitive exclusion~\cite{Gause:1960}. 

In contrast to that previous work, in this paper we address the heterogeneity of networks. Networks can differ in functionality, features, and --most importantly-- they can address different peer groups. Here, we show how the effect of different overlapping peer groups can be described in terms of different degrees of network fitness. We find that under certain conditions, the heterogeneity of degrees of fitness can impede coexistence which would indeed be possible for identical networks. This effect is particularly important for the competition between local networks and an international network. 
Unlike users of local networks, users of the international network have the possibility to interact with people in other countries, providing this network an advantage over local ones, similar to a higher fitness of a certain species. A proper modeling of this effect requires taking into account the network of interactions among countries in the world, which results in a highly complex and non-linear dynamical system made of interconnected multilayer networked entities -- we are hence dealing with networks of multiplex networks~\cite{Reis2014,Radicchi2014,Bianconi2014,DAgostino2014:Netonets2013}, in contrast to Ref.~\cite{ecology20}. Besides, inter-country interactions induce a different type of bifurcation as the symmetry of the previous model is broken, which constitutes a fundamentally new behavior not observed in~\cite{ecology20}.

Empirical observations have shown that Facebook expanded massively
in the middle of the first decade of this century, starting in the US, when local networks were the most popular services in most countries. 
Only a few years later, Facebook had become the most popular network in most countries. So, is the fate of the digital world to become dominated by a single ``big brother'' as it takes over all our digital interactions? Alternatively, is digital diversity possible from a system-level perspective? In this paper, 
we show that due to the nonlinear character of 
our model, the answer to both questions can be positive or negative depending on a range of parameters and, quite surprisingly, depending on chance. 
As we will show, our model, despite its simplicity and the limited number of parameters, is able to describe surprisingly well the complex behavior of globally interacting online social networks.

\section{Results}

\begin{figure*}[t]
 \includegraphics[width=1\linewidth]{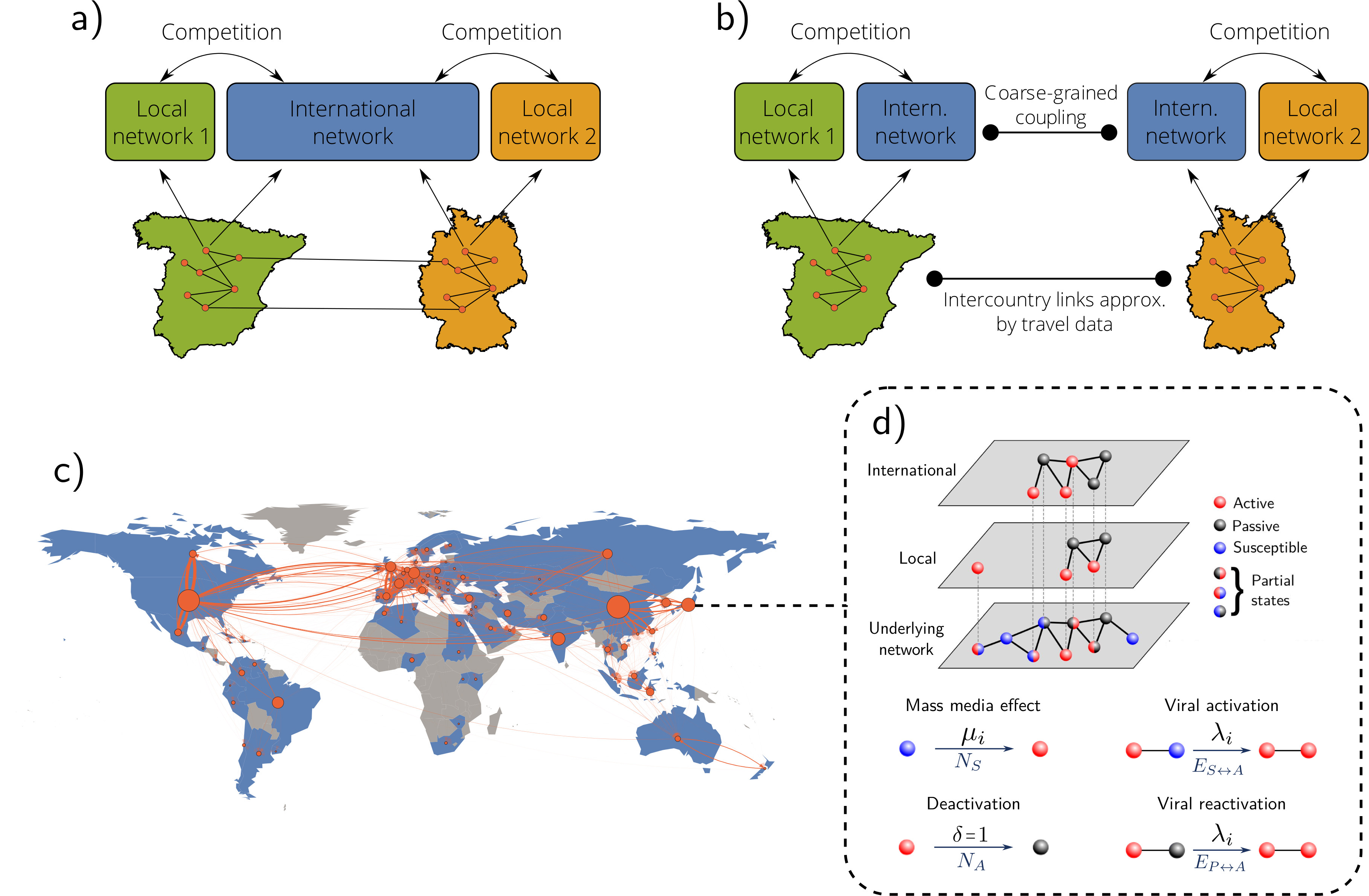}
 \caption{Constituents of our model. \textbf{a)} Design of the international network and local networks. \textbf{b)} Sketch of our model using coarse-grained coupling. \textbf{c)} Visualization of the flight network. The size of the nodes is proportional to the number of users in the respective countries with Internet access. The transparency and thickness of the links represents the density of passengers between the countries concerned. \textbf{d)} Illustration of the competition between the international network and the local network within one country.
 \label{fig_panel1}}
\end{figure*}

\subsection{Complex organization of the digital world}
\label{sec_sum_evo_eco}

The digital world consists of highly connected and strongly coupled interacting subsystems. These basic building blocks are single networks, each of which obeys specific dynamics in the absence of coupling to the whole system. 
So the complexity of the digital world is a consequence of both the dynamics of networks in isolated environments and the interactions between many such networks. Finally, not all of these building blocks are identical. Instead,
different networks address different peer groups or have different functionalities.
Hence, to reveal the fundamental mechanisms that determine the fate of the digital world, it is necessary to understand the interaction of heterogeneous networks, each driven by intrinsic dynamics.

\subsection{Isolated dynamics of online social networks}

The key actors in
the digital world are OSNs; loosely defined as web-based platforms that enable digital social interactions over the Internet.
However, societies were organized as networks long before OSNs were even thought of. 
From this point of view, the growth of OSNs can be described through the dynamical processes by which people in the traditional offline social structure come to engage in OSNs. The topology of the OSN is now the digital counterpart of the underlying offline social network~\cite{our:model,clauset:facebook}.

In isolation, this process of formation 
can be described by a set of simple dynamical mechanisms~\cite{our:model}. The system is initially given by an empty OSN and the underlying social structure. Individuals can be in three different states: active, passive or susceptible. While active and passive nodes exist in the online as well as the offline networks, susceptible nodes are only present in the latter. A susceptible node can join the OSN via two different mechanisms: a \Hl{viral activation} effect which means that a susceptible node becomes active due to the influence of an active neighbor in the offline network and a \Hl{mass media} effect, which represents the spontaneous activation of a susceptible node. In addition, an active node can become passive spontaneously (\Hl{deactivation}) and a passive node can become active again due to the influence of an active neighbor (\Hl{viral reactivation}). See Fig.~\ref{fig_panel1}d for a visualization of these mechanisms. Notice that, 
in the long time limit, when the number of susceptible individuals is basically zero, this dynamics is equivalent to the susceptible-infected-susceptible model widely used in epidemiology~\cite{anderson1992infectious}.
As found in\cite{our:model,Ribeiro:www:2014},
this implies that online social networks can either exhibit a sustained activity (similar to the endemic phase), or they can become entirely passive (similar to the healthy phase).

The evolution of OSNs rarely takes place in isolation. Nevertheless, we found 
a perfect case study in the Slovakian social network ``Pokec'', which, due to the particularities of the country, has been growing in quasi-isolation for more than ten years. By analyzing the evolution of the topology of the social contact graph of ``Pokec'', 
we were able to rigorously validate our model. 
Quite remarkably, with only two parameters, the model reproduced the entire topological evolution with astonishing precision.

\Hl{Viral activation} and \Hl{viral reactivation} occur at the same rate, $\lambda$, and the ratio between this rate and the rate of \Hl{mass media} influence, $\mu$, governs the topological evolution. 
In particular, we observed that the real system underwent a dynamical percolation transition; that is, a phase transition between a disconnected phase and a phase in which a macroscopic fraction of the system is connected. The position of this transition is controlled by $\lambda / \mu$, due to the complementary roles that the \Hl{viral activation} and {\Hl{mass media}} effects play in the topological evolution of the network (the former tends to connect components; whereas the latter tends to create new components).
Finally, without loss of generality, we set the \Hl{deactivation} rate $\delta$ to $1$ which 
is equivalent of fixing the timescale of the model. 

To sum up, in our previous work~\cite{our:model}, we were able to rigorously validate the dynamics ruling OSNs in isolation; the fundamental building blocks of the digital world. 
These findings constitute the foundation for the development of a more comprehensive theory of interacting heterogeneous networks.

\subsection{Competitive interaction between multiple networks}

The simultaneous existence of multiple digital services in competition for the attention of users suggests an ecological perspective from which to explain the prevalence of a given network or the coexistence of multiple networks. In ecology theory, the principle of competitive exclusion~\cite{Gause:1960} states that multiple species in competition for the same resource cannot coexist, as even the slightest advantage of one species over the other is successively amplified; a mechanism referred to as ``rich get richer'' or preferential attachment~\cite{Barabasi:1999ha,DoMeSa01,BiBa01a,CaCaDeMu02,Vazquez2003,PaSaSo03,FoFl06,SoBo07,boguna:popularity}.
This eventually leads to the extinction of the inferior species. 

The key principle that drives the competition between OSNs is the fact that, due to the physical and cognitive limitations of users, the time they devote to online activities is limited. As a consequence, the viral parameter, $\lambda$, constitutes a conserved quantity that is nevertheless distributed between the competing networks as $\lambda_i = \omega_i(\gvec{\rho}^\mathrm{a}) \lambda$, where $\omega_i(\gvec{\rho}^\mathrm{a})$ represents a normalized set of weights, that is $\sum_i \omega_i(\gvec{\rho}^\mathrm{a}) = 1$, and $\gvec{\rho}^\mathrm{a}  \equiv (\rho_1^\mathrm{a},\rho_2^\mathrm{a},\dots)$ is a vector denoting the fraction of active nodes (activities) in the different networks.
In general, users are more likely to subscribe to and engage in networks that are more active. 
Therefore, the viral activity of each network must be a function of the activity of the network itself. In particular, we model this by assuming that the weighting $\omega_i(\gvec{\rho}^\mathrm{a})$ is a function, such that
$\partial \omega_i(\gvec{\rho}^\mathrm{a}) / \partial \rho_i^\mathrm{a} >0$. 
In~\cite{ecology20} we proposed the particular form:
\begin{equation}
 \omega_i(\gvec{\rho}^\mathrm{a}) = \frac{\left[ \rho_i^\mathrm{a} \right]^\sigma}{\sum_{j=1}^{n_l} \left[ \rho_j^\mathrm{a} \right]^\sigma} \eqcomma
 \label{eq_weights}
\end{equation}
were $n_l$ denotes the number of networks. 
This choice allows us to interpolate between a set of independent networks ($\sigma \ll 1$) and highly coupled ones ($\sigma \gg 1$).
The activity affinity parameter, $\sigma$, then quantifies the tendency of users to subscribe to or engage in more active networks.
Interestingly, in contrast to the principle of competitive exclusion, multiple networks can coexist because the ``rich get richer'' mechanism is damped by the diminishing returns of the dynamics of network evolution. 
For details, we refer the reader to Ref.~\cite{ecology20}. 
In the following, we take into account the heterogeneity of networks induced by different groups of individuals that can subscribe to the different networks. These aspects are not discussed in Ref.~\cite{ecology20} and have important implications and applications, as we will show.

\subsection{Network heterogeneity leads to effective activity}

\label{sec_effective_activity}

As mentioned above, since its official launch in 2004, Facebook has become the most popular OSN in most countries; even in countries where there was already a popular OSN before Facebook was launched. 
To mimic the real evolution of the digital ecosystem at the worldwide scale, we assume that one local network exists in each country in addition to a globally operating, international network (see Fig.~\ref{fig_panel1}a).
In the US, both networks are launched at the same time; whereas the international network is launched with a delay $\Delta t$ in the remaining countries, to take into account the initial prevalence of local networks. 

Once launched, the international network provides the user with the possibility to connect to individuals in different countries, in contrast to local networks, making it more attractive to users. 
For a given country, the advantage of the international network is directly related to the abundance of social ties between that country and the rest of the world. 
We use passenger air travel data as a proxy for the abundance of such ties. This choice is justified by the strong correlation between air travel flows and further measures of inter-country exchange, for instance email communication~\cite{macy:mail} or Twitter activity~\cite{twitter:geo}. 

Users in country $i$ experience the greater attractiveness of the international network 
as 
they perceive its activity with respect to the population of their own country and also with respect to their contacts in other countries. To account for this on a coarse grained level, 
in Eq.~\eqref{eq_weights}, we replace the activity of the international network by an effective activity as follows

\begin{equation}
  \tilde{\rho}_{i,\text{int}}^{\mathrm{a}} = \rho_{i,\text{int}}^{\mathrm{a}} +  \alpha \sum_{j} \Omega_{ij} \rho_{j,\text{int}}^{\mathrm{a}} \eqcomma
  \label{eqn_effective_activity}
\end{equation}
where \begin{equation}
       \Omega_{ij} = \frac{W_{ij} / N_i}{\max[W_{ij} / N_i]} 
       \label{eqn_def_omega}
      \end{equation}
denotes the fraction of the number of air travel passengers between countries $i$ and $j$, $W_{ij}$; 
and $N_i$, the population of country $i$.
Notice that, in an ecological context, this corresponds 
to increased fitness of the international network.
In Eq.~\eqref{eqn_effective_activity}, we have implicitly assumed proportionality between the number of passengers and the number of contacts in the respective countries, namely $N_{ij} \propto W_{ij}$. 
Finally, note that the arbitrary normalization in Eq.~\eqref{eqn_def_omega} serves the sole purpose of ensuring that reasonable values for the parameter $\alpha$ are of the order of unity. 

Hereafter, we decompose the international network into a set of disjunct coupled subnetworks operating in each country and in competition with the respective local network (see Fig.~\ref{fig_panel1}b). 
These subnetworks are nevertheless not independent, as they are globally coupled via the effective activity defined in Eq.~\eqref{eqn_effective_activity} and ultimately by the network representing the inter-country social ties. Hence, our model forms a network of networks~\cite{Reis2014,Radicchi2014,Bianconi2014,DAgostino2014:Netonets2013}, where each node in Fig.~\ref{fig_panel1}c represents a three-layer multiplex network~\cite{arenas:multiplex,multilayer:kivel} in which the bottom layer corresponds to the underlying social structure and the two upper layers denote the local and international networks operating in the respective country (see Fig.~\ref{fig_panel1}d).

\subsection{Double meanfield approximation reveals complex role of the activity affinity}
\label{sec_meanfield}
\begin{figure*}[p]
\includegraphics[width=1\linewidth]{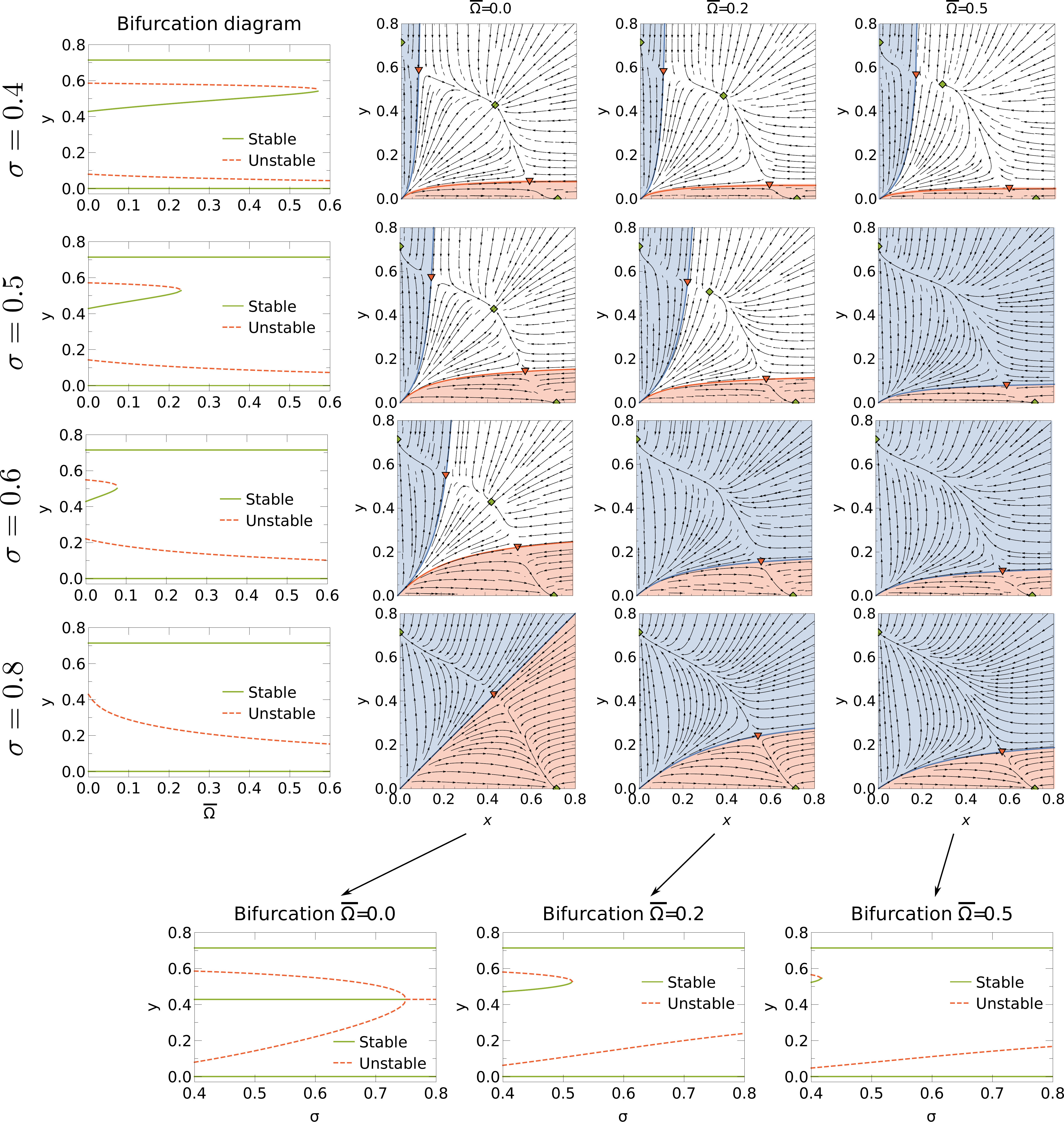}
 \caption{Bifurcation diagram  and stream plots for the double meanfield approximation~\eqref{eqn_doublemf} for $\lambda \km = 3.5$ and $\nu \rightarrow \infty$.
 \label{fig_bif} \label{fig_stream}
The basins of attraction for the domination of the international network are in blue, the basins of attraction for the domination of local networks are in red; the white areas correspond to the basins of attraction of the coexistence solution (if it exists).}
\end{figure*}

To understand the qualitative behavior of the system, in this section
we present a double meanfield approximation of the system. 
This reduces the system given by a network of networks to a set of evolution equations of the average activity in the international network and in local networks. 
As we show 
in the following section, the results of the full model with heterogeneous topologies exhibits similar behavior to that encountered by the double meanfield approximation.

The first meanfield approximation consists of assuming a fully mixed homogeneous population in each country. 
Let $\rho^\mathrm{a}_{i,l}$ denote the fraction of active users in network $l \in (\text{loc},\text{int})$ in country $i$ and $\rho^\mathrm{s}_{i,l}$ the fraction of nodes susceptible to joining this network. Then, the fraction of passive users is given by $1-\rho^\mathrm{s}_{i,l}-\rho^\mathrm{a}_{i,l}$.
As explained above, in each country the virality is distributed between the local and international network via the weight functions
$
 \omega_{\text{loc}}(\rho_{i,\text{loc}}^\mathrm{a},\tilde{\rho}_{i,\text{int}}^\mathrm{a})$ and
$ \omega_{\text{int}}(\rho_{i,\text{loc}}^\mathrm{a},\tilde{\rho}_{i,\text{int}}^\mathrm{a}) = 1 - \omega_{\text{loc}}(\rho_{i,\text{loc}}^\mathrm{a},\tilde{\rho}_{i,\text{int}}^\mathrm{a})$, 
as introduced in Eq.~\eqref{eq_weights}.
Here, $\tilde{\rho}_{i,\text{int}}^\mathrm{a}$ denotes the effective activity of the international network as defined in Eq.~\eqref{eqn_effective_activity}. The evolution equations of the resulting system represent a generalization of the evolution equations for identical networks which we derived in~\cite{ecology20}, where one replaces the activity of the international network with the effective activity from Eq.~\eqref{eqn_effective_activity}. This procedure yields:
\begin{align}
 \begin{split}
 \dot{\rho}^\mathrm{a}_{i,l} & = \rho^\mathrm{a}_{i,l} \biggr\{ \lambda \km \omega_{l}(\rho_{i,\text{loc}}^\mathrm{a},\tilde{\rho}_{i,\text{int}}^\mathrm{a})  \left[ 1 - 
\rho^\mathrm{a}_{i,l} \right] -1 \biggl\}  + \mu_{i,l} \rho_{i,l}^\mathrm{s}\\
\dot{\rho}_{i,l}^\mathrm{s}& = -\mu_{i,l} \rho_{i,l}^\mathrm{s} \biggr\{1+\nu \km \rho_{i,l}^\mathrm{a}   \biggl\}
\eqdot
\label{eq_meanfield_full}
 \end{split}
\end{align}
As shown in~\cite{our:model,ecology20}, we further assume the the same linear relationship between virality and media influence in each country, that is:
\begin{align}
\begin{split}
 \mu_{i,\text{loc}} &= \frac{\lambda \omega_{\text{loc}}(\rho_{i,\text{loc}}^\mathrm{a},\tilde{\rho}_{i,\text{int}}^\mathrm{a})}{ \nu} \\
 \mu_{i,\text{int}} &= \frac{\lambda \omega_{\text{int}}(\rho_{i,\text{loc}}^\mathrm{a},\tilde{\rho}_{i,\text{int}}^\mathrm{a})}{ \nu}
 \eqdot
 \label{eqn_introduce_nu}
 \end{split}
\end{align}
As shown in~\cite{ecology20}, the value of $\nu$ does not affect the stability of the system. In what follows, we perform the stability analysis in the limit $\nu \rightarrow \infty$. This decouples the evolution of $\rho^\mathrm{a}_{i,l}$ from $\rho^\mathrm{s}_{i,l}$, so that we only have to consider $\rho^\mathrm{a}_{i,l}$.
Plugging in the weights function defined in Eq.~\eqref{eq_weights} and the effective activity from Eq.~\eqref{eqn_effective_activity} yields the evolution equations for the activities of the local and international networks in country $i$:
\begin{align}
 \begin{split}
  \dot{\rho}_{i,\text{loc}}^{\mathrm{a}} & = \rho_{i,\text{loc}}^{\mathrm{a}} \left[  \frac{\lambda \km [\rho_{i,\text{loc}}^{\mathrm{a}}]^\sigma}{[\rho_{i,\text{loc}}^{\mathrm{a}}]^\sigma + (\rho_{i,\text{int}}^{\mathrm{a}}+\delta_i)^\sigma} [1-\rho_{i,\text{loc}}^{\mathrm{a}}]-1\right] \\
    \dot{\rho}_{i,\text{int}}^{\mathrm{a}} & = \rho_{i,\text{int}}^{\mathrm{a}} \left[ \frac{\lambda \km  (\rho_{i,\text{int}}^{\mathrm{a}}+\delta_i)^\sigma}{[\rho_{i,\text{loc}}^{\mathrm{a}}]^\sigma + (\rho_{i,\text{int}}^{\mathrm{a}}+\delta_i)^\sigma} [1-\rho_{i,\text{int}}^{\mathrm{a}}]-1\right] \eqcomma
 \end{split}
 \label{eqn_meanfield}
\end{align}
where $\delta_i = \alpha \sum_{j} \Omega_{ij} \rho_{j,\text{int}}^{\mathrm{a}}$.

The second meanfield approximation consists of applying the hypothesis of a fully mixed homogeneous network for the inter-country social ties. We use $\bar{\Omega} = \alpha \left< \Omega_{ij} \right>$ and define the mean activity of the local networks as $x \equiv \langle \rho_{i,\text{loc}}^{\mathrm{a}} \rangle$ and the mean activity of the international network as $y \equiv \langle \rho_{i,\text{int}}^{\mathrm{a}} \rangle$. Finally, our double meanfield approximation leads to the following system of coupled differential equations
\begin{align}
 \begin{split}
  \dot{x} & = x \left[ \lambda \km \frac{x^\sigma}{x^\sigma + (y (1+\bar{\Omega}))^\sigma} [1-x] -1 \right] \\
  \dot{y} & = y \left[ \lambda \km \frac{(y (1+\bar{\Omega}))^\sigma}{x^\sigma + (y (1+\bar{\Omega}))^\sigma} [1-y]  -1 \right] \eqcomma
  \label{eqn_doublemf}
 \end{split}
\end{align}
which has three relevant 
parameters: $\lambda \km$, $\sigma$, and $\bar{\Omega}$. Note that by setting $\bar{\Omega}$ to zero, we recover the equations for identical networks presented in~\cite{ecology20}.

In what follows, we discuss the dynamical properties of the system given by Eq.~\eqref{eqn_doublemf}.
For constant $\sigma$, the system exhibits a saddle-node bifurcation at a critical value of the global connectivity $\bar{\Omega}_c(\sigma)$ (see Fig.~\ref{fig_bif}). Above this point, coexistence is not possible and the only stable solutions correspond to the domination of either local networks or the international one. 
Both above and below the critical value 
$\bar{\Omega}_c(\sigma)$, the basin of attraction of the solution corresponding to the domination of local networks decreases with $\bar{\Omega}$, whereas that of the international network increases (see the rows of Fig.~\ref{fig_bif}).
Furthermore, at the critical point, the basin of attraction of the international network is amplified discontinuously as the region of coexistence in the subcritical regime is now merged with the basin of attraction of the domination of the international network.

For constant $\bar{\Omega}>0$, the system also exhibits a saddlenode bifurcation
at a critical value of the activity affinity $\sigma_c(\bar{\Omega})$. 
In~\cite{ecology20} we showed that the system undergoes a subcritical pitchfork bifurcation with respect to the control parameter $\sigma$, 
above which no stable coexistence is possible. $\bar{\Omega} > 0$ breaks the symmetry of the pitchfork bifurcation and in this case the system undergoes a saddlenode bifurcation with respect to $\sigma$ instead (see bottom of Fig.~\ref{fig_bif}). This behavior is well known in bifurcation theory and results from adding a small error term to the normal form of the pitchfork bifurcation (see Supplementary Materials).
The evolution of the basins of attraction is more complex compared to the previous case. Below the critical point, both basins of attraction increase with $\sigma$. Above the critical point $\sigma_c(\bar{\Omega})$, the basin of attraction of the local network increases whereas the basin of attraction of the international network decreases with $\sigma$ (see the columns of Fig.~\ref{fig_bif}). This is particularly interesting as it implies that an intermediate value of the activity affinity just slightly above the critical point $\sigma \gtrsim \sigma_c(\bar{\Omega})$ represents the worst scenario for the survival of local networks, since at this point the size of the basin of attraction of the domination of the international network is maximum.

\begin{figure}[t]
\centering
 \includegraphics[width=1\linewidth]{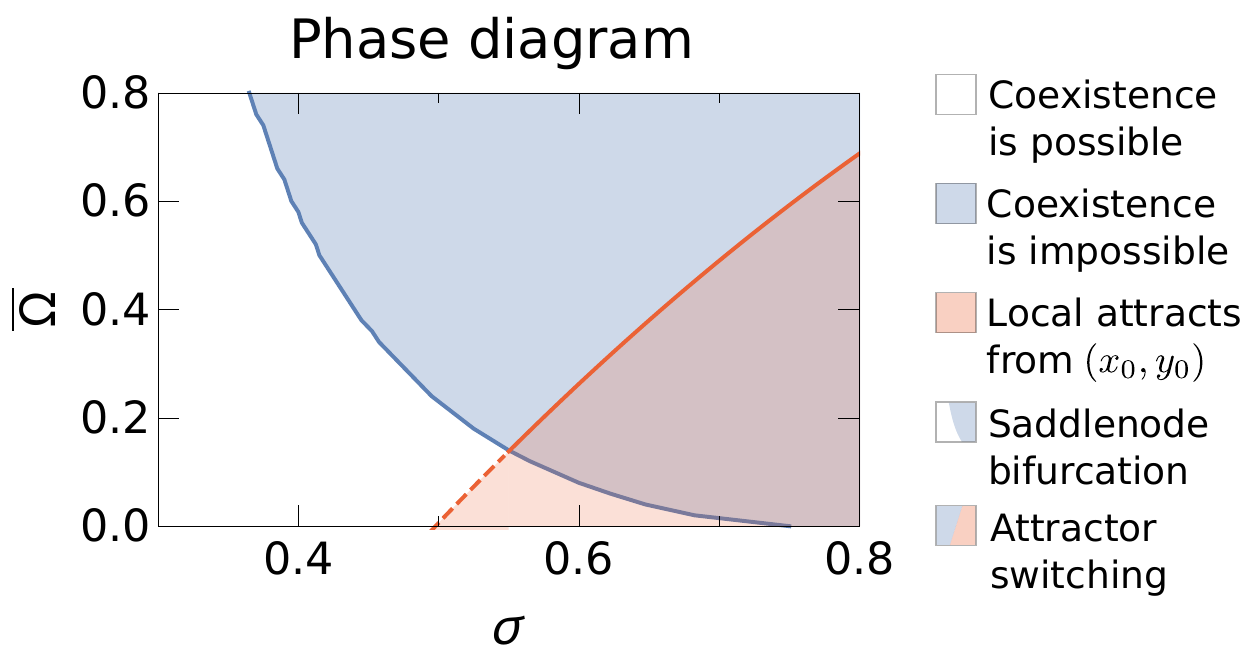}
 \caption{Phase diagram of the double meanfield approximation for $\lambda \km = 3.5$.  
The white area denotes the parameters for which coexistence is possible. 
The blue area denotes the parameters for which only domination of either local or the international network can occur. 
At the blue line, the system undergoes a saddle-node bifurcation in which the stable coexistence solution disappears. The red line denotes the combination of parameters for which the system switches attractors for the initial conditions given by Eq.~\eqref{eqn_initial}, where we use $\beta=0.2$ to reflect that the US contributes about $20\%$ of the population.
The red region shows the parameters for which the solution of domination of local networks is reached for these initial conditions.
Above the red line, the system approaches the domination of the international network (solid red line) or the coexistence solution (dashed red line).
 \label{fig_phase}}
\end{figure}

In Fig.~\ref{fig_phase}, the blue line indicates the critical line $\bar{\Omega}_c(\sigma)$ in the $\sigma$-$\bar{\Omega}$ plane, which separates a phase in the parameter space where coexistence is possible (white region) and one in which only domination can occur (blue region). 
However, the increasing size of the basin of attraction of the domination of local networks above the critical point with respect to $\sigma > \sigma_c(\bar{\Omega})$ can dramatically alter the fate of the system for a given set of initial conditions. 
Assume, for instance, that the international network dominates in the US and starts with a significant delay in each other country, which causes the local networks to dominate in those countries.
At the time when the international network is launched globally, the state of the system can be approximated as follows
\begin{align}
 \begin{split}
  x_0 & = (1-\beta) \left[1-\frac{1}{\lambda \km}\right] \\
  y_0 & = \beta \left[1-\frac{1}{\lambda \km}\right] \eqcomma
  \label{eqn_initial}
 \end{split}
\end{align}
which we now use as initial conditions to study the further evolution. 
Notice that if one network dominates in country $i$, its activity is given by $\rho_{i,l}^\mathrm{a} = 1-\frac{1}{\lambda \km}$.
Hence, the initial conditions given by Eq.~\eqref{eqn_initial} reflect the fact that local networks dominate in the fraction $(1-\beta)$ of the system and the international one dominates in the remainder.
The evolution of the basins of attraction makes the system approach different stationary solutions from these initial conditions for different parameters. 
Below the red line in Fig.~\ref{fig_phase}, the system approaches the domination of local networks starting from the initial conditions given in Eq.~\eqref{eqn_initial}. Above this line, the system either approaches coexistence (white area; crossing dashed red line) or domination of the international network (blue area; crossing solid red line). This means that in the red region, when the international network is launched globally, it is not able to overcome the initial advantage of the local networks due to its earlier launch.

To conclude, the double meanfield approximation predicts that intermediate values of the activity affinity most favor the international network; whereas the local networks can dominate for a high activity affinity and low global connectivity. We confirm these findings by numerical simulations in the following section.

\subsection{Numerical simulations and synthetic networks}
\label{sec_scale_model}

\begin{figure*}[t]
 \includegraphics[width=1\linewidth]{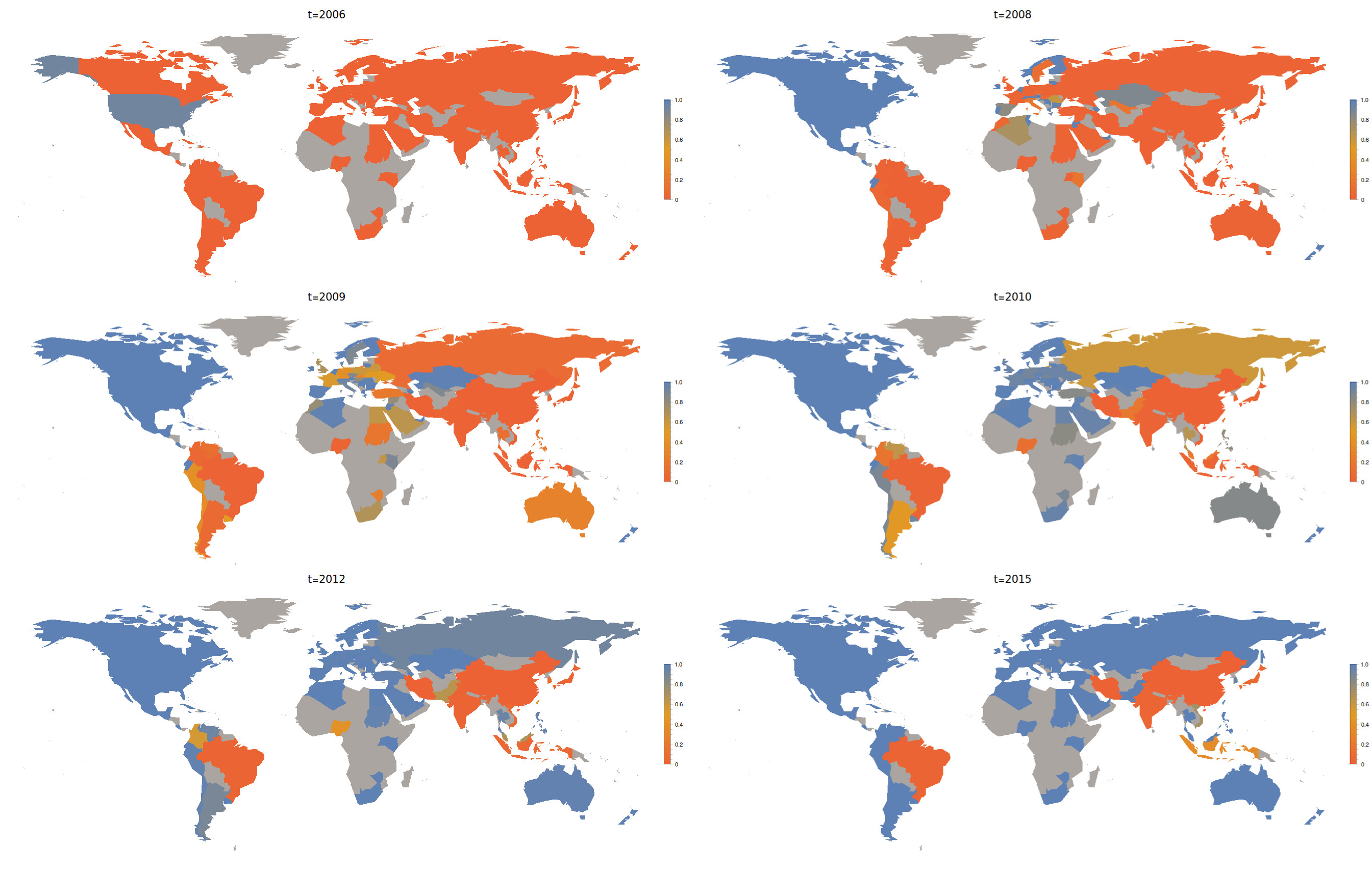}
 \caption{
Evolution of our model averaged over several realizations for the parameters
 $\sigma = 1.25$, $\Delta t = 3$, and $\alpha = 2$ (here, we excluded realizations in which local networks dominate, which occurs with approximately $30\%$ probability for these parameters). The mapping from model time to real time is explained in the following section. 
 The relative importance of the influence of mass media compared to the viral spreading mechanism in governed by the parameter $\nu$ introduced in Eq.~\eqref{eqn_introduce_nu}. In all numerical simulations, we set $\nu = 4$: the value found empirically in~\cite{our:model}; and $\lambda = 0.2$ (this corresponds to $\lambda / \lambda_c^1 \approx 4.3$ in~\cite{ecology20}).
 The relative prevalence of the international network, given by $\rho_{i,\text{int}}^{\mathrm{a}}/(\rho_{i,\text{int}}^{\mathrm{a}}+\rho_{i,\text{loc}}^{\mathrm{a}})$, is color coded. 
  We consider the international network to be banned in China and Iran. To model this, we set the values of $\Omega_{ij} = 0$ for each entry which involves one of these countries. This is equivalent to assuming that in these countries two local networks compete without any coupling to the rest of the world.
  \label{fig_w8}}
\end{figure*}

In this section, we go beyond the meanfield approximation and study, by means of numerical simulations, the effects of the real topology of inter-country social ties and of underlying social structures.
To this end, 
we use the air travel network (see Fig.~\ref{fig_panel1}c and Materials and Methods) as a proxy for inter-country social ties and
construct 1:1000 scaled synthetic networks to model the structure of the $80$ countries with most Internet users (see Tab.~\ref{tab_countries}). To generate these networks, we make use of a model introduced in~\cite{Serrano2008,Boguna2008,Boguna2003}, which produces realistic topologies of the traditional offline social networks, including heterogeneous node degrees and a high level of clustering (see Materials and Methods).

Fig.~\ref{fig_w8} shows results from our model for the set of parameters that best matches empirical observations, as explained in the following section.
The international network starts with a delay in all countries except the US; so that initially in these countries the respective local network dominates. After some time, the international network obtains a significant advantage and quickly takes over in most countries.

\begin{figure*}[t]
\centering
 \includegraphics[width=0.8\linewidth]{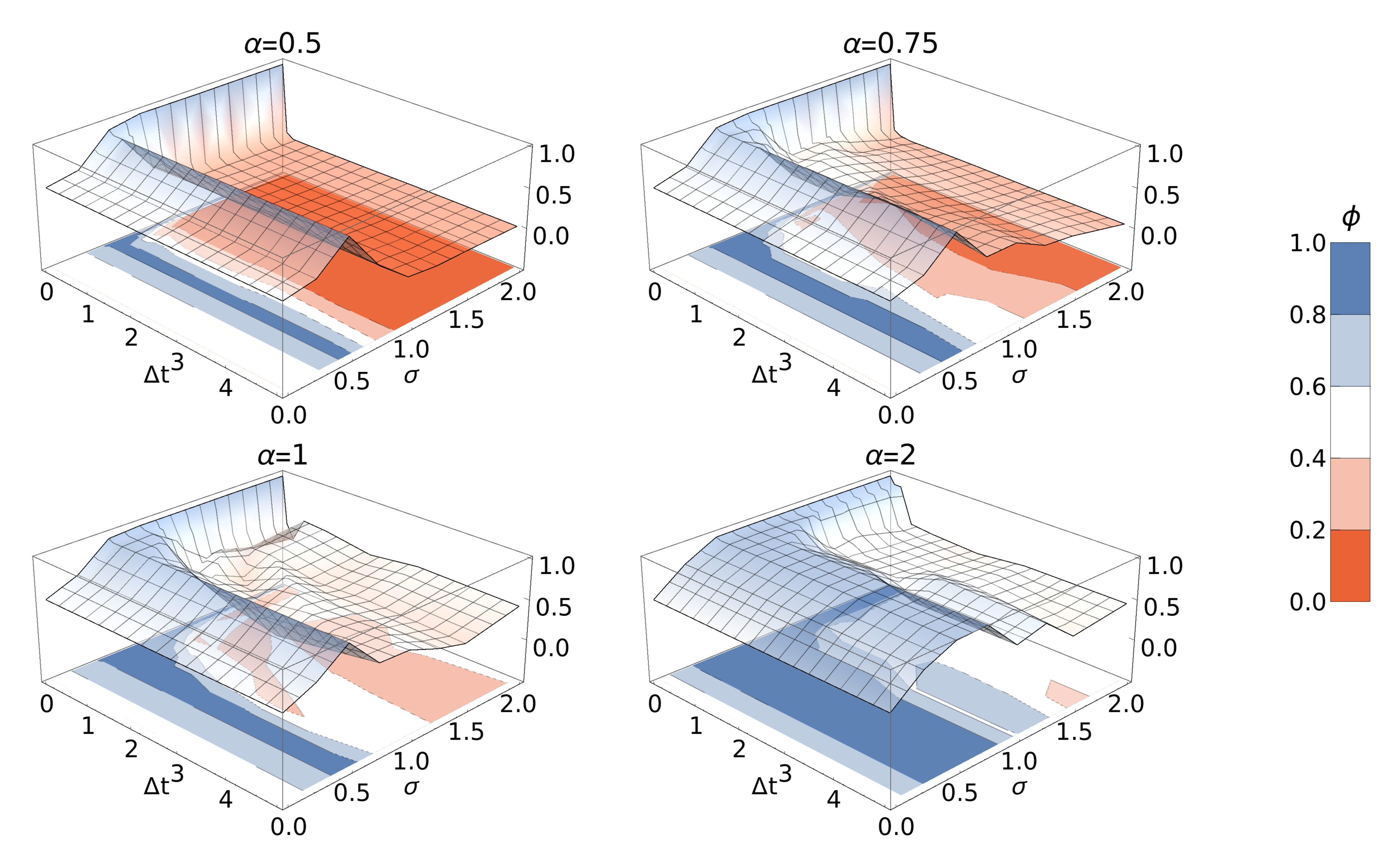}
 \caption{Relative prevalence $\Phi$ of the international network is plotted on the $z$ axis as a function of the launch time delay $\Delta t$ and the coupling strength $\sigma$. Averaged over 30 realizations. \label{fig_numerical} \label{fig_density}}
\end{figure*}
\begin{figure*}[t]
\centering
\includegraphics[width=1\linewidth]{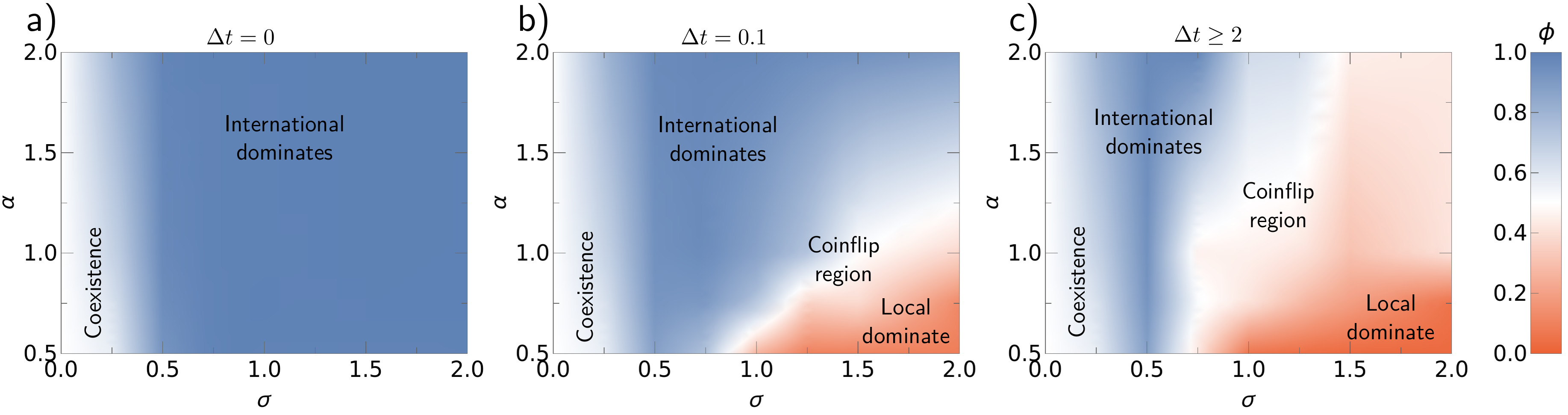}
 \caption{
The prevalence of the international network for \textbf{a)} $\Delta t = 0$, \textbf{b)} $\Delta t = 0.1$, and \textbf{c)} averaged over time delays $\Delta t \geq 2$ as a function of the activity affinity ($\sigma$) and the global connectivity ($\alpha$). Averaged over 30 realizations. 
 \label{fig_tcut}}
\end{figure*}

To further study the properties of the model presented here, we define the relative prevalence of the international network compared to local networks as:
\begin{equation}
 \Phi =\frac{1}{n_c} \sum_{i=1}^{n_c} \frac{  \stst{\rho_{i,\text{int}}^{\mathrm{a}}} } { \stst{\rho_{i,\text{int}}^{\mathrm{a}}}  + \stst{\rho_{i,\text{loc}}^{\mathrm{a}}}  }
\end{equation}
where $n_c$ denotes the number of countries, and $\stst{\rho_{i,\text{int}}^{\mathrm{a}}}$ and $\stst{\rho_{i,\text{loc}}^{\mathrm{a}}}$ are the activities of the international and local networks in country $i$ in the stationary state.
With this definition, a value of $\Phi \approx 0$ implies that local networks dominate in most countries, whereas $\Phi \approx 1$ corresponds to the domination of the international network. The relative prevalence of the international network averaged over many realizations is shown in Fig.~\ref{fig_density} for different values of $\alpha$ as a function of the activity affinity, $\sigma$, and the launch time delay, $\Delta t$. 
For small values of $\Delta t$, we observe that when $\sigma$ is small, the international and local networks coexist and we observe values around $\Phi \approx 0.5$ for the relative prevalence; then, increasing $\sigma$ favors the international network, which dominates for values of $\sigma \gtrsim 0.5$ (see Fig.~\ref{fig_tcut}a and b). For larger values of $\Delta t$, this behavior smoothly translates into a more complex case, which we discuss below.

\begin{figure*}[t]
\centering
 \includegraphics[width=1\linewidth]{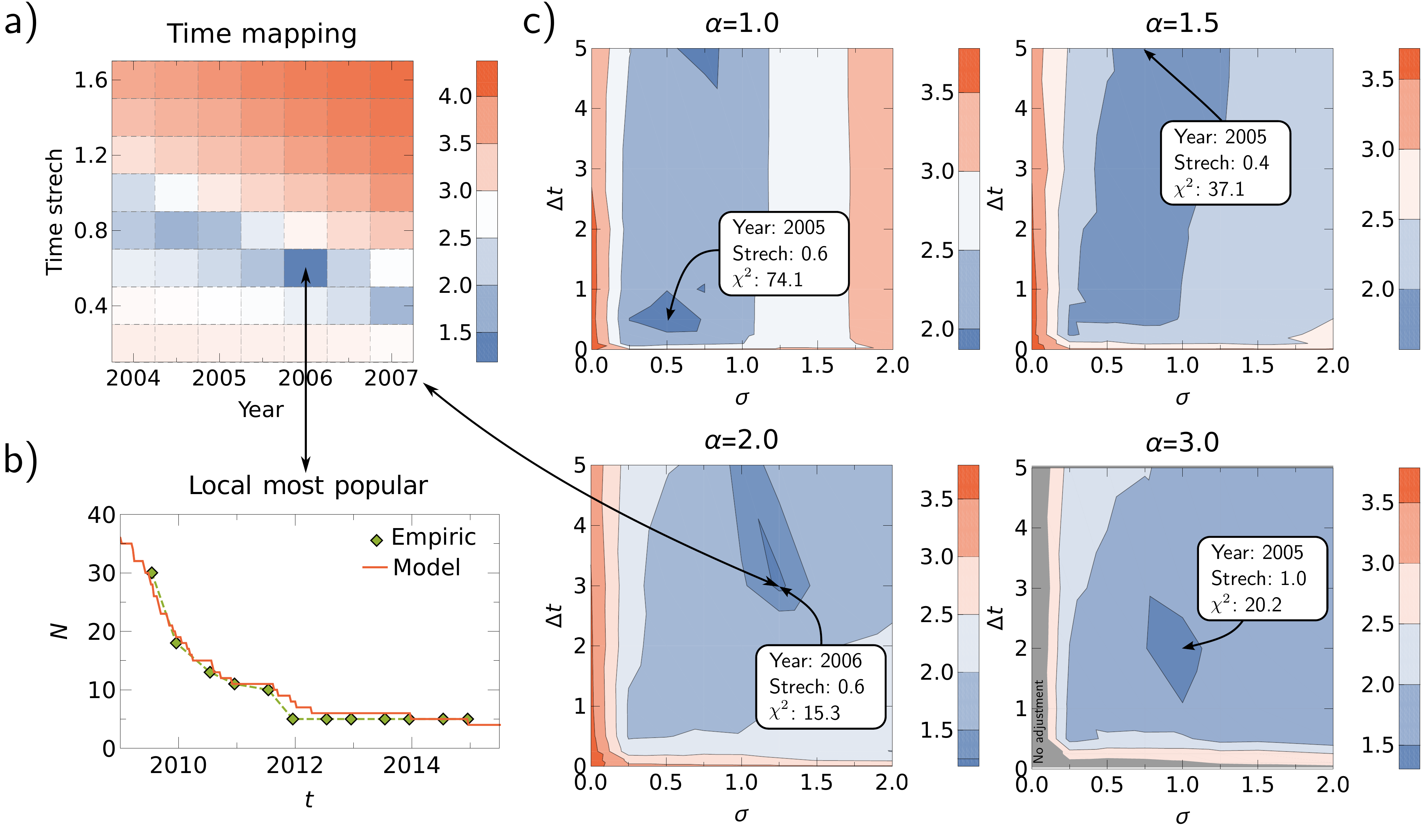}
 \caption{
 Comparison between model results and empirical data. \textbf{a)} $\chi^2$ for different values of $a_1$ (year) and $a_2$ (time stretch) for $\sigma=1.25$, $\alpha=2$, and $\Delta t = 3$. \textbf{b)} Model results and empirical data. Here, the parameters are $\sigma=1.25$, $\alpha=2$, and $\Delta t = 3$ and the optimal mapping is given by $a_1=2006$ and $a_2=0.6$ (see \textbf{a)}. \textbf{c)} $\chi^2$ for the respective best time mapping as a function of $\sigma$ and $\Delta t$ for different values of $\alpha$. In each plot, the minimal value of $\chi^2$ and the respective time mapping are shown in the boxes. The color coding in all plots represents the logarithm of $\chi^2$.
 \label{fig_compare_empiric}}
\end{figure*}

We observe in Fig.~\ref{fig_density} that for launch time delays $\Delta t \geq 2$, the actual length of the delay becomes irrelevant. This behavior corresponds to the limit of saturation of the evolution of local networks before the international OSN is launched; 
as discussed 
% \sout{is section} 
in the previous section.
We consider this limit by averaging over regions with $\Delta t \geq 2$ in Fig.~\ref{fig_tcut}c, which yields a two dimensional parameter space $\sigma$-$\alpha$.
Indeed, numerical simulations of the full model confirm the results from the meanfield analysis; in particular the complex role of the activity affinity $\sigma$. 
For small $\alpha$ and $\sigma$, local networks and the international OSN can coexist. Increasing $\sigma$ or $\alpha$ favors the domination of the international network, which gives rise to the blue ``V''-shaped region around $\sigma=0.5$. This corroborates the saddlenode bifurcation predicted by the double meanfield approximation. 
See supplementary video for an explicit realization.
For high values of $\sigma$ and small values of $\alpha$ (red region in the  bottom right-hand corner of Fig.~\ref{fig_tcut}), local networks dominate.  
Note that partial states are also possible, in which the international network dominates in some countries and local networks dominate in the remaining countries. See supplementary video for an explicit realization of this case.

Between the regions of domination of the international network and of local networks, there is a region in which the final fate of the system varies significantly between different realizations of the model (``coinflip region''). In this region, 
if the international network wins initially in the US, it will become dominant globally; otherwise, local networks maintain their initial prevalence.
Although in this region the prevalence of the international network averaged over many realizations is about $0.5$, as in the coexistence region in the bottom left-hand corner of Fig.~\ref{fig_tcut}, the behavior of the system differs dramatically from one to another. In the coexistence region, each realization of the model leads to the same final state: coexistence of local networks and the international OSN. In contrast, in the coinflip region, coexistence is not possible, as this region of the parameter space corresponds to the supercritical regime (the blue area in Fig.~\ref{fig_phase}). In the coinflip region, about $50\%$ of the realizations end up with domination of the international network, whereas the remaining $50\%$ lead to the domination of local OSNs. As a consequence, even if we know the exact parameters, it is impossible to predict the fate of the system beforehand. 

We can summarize these findings as follows. A higher value of $\alpha$, which is a measure of the global connectivity of society, favors the prevalence of the international network and hinders the survival of the local ones. The role of the tendency of individuals to participate in more active networks (activity affinity), $\sigma$, is particularly interesting. Low values allow the networks to coexist, whereas intermediate values always lead to the prevalence of the international network and the extinction of local OSNs. A high activity affinity, however, enables the prevalence of local networks and thus can even lead to the extinction of the international network.

\subsection{Comparison with empirical data}

\label{sec_emp_data}

In this section, we compare the results of our model with empirical data on the recent expansion of Facebook at the cost of many local networks. In particular, we consider the evolution of the number of countries in which local networks (i.e. networks that are not Facebook) are the most popular ones, as measured in~\cite{worldmap} using Alexa traffic data (see Fig.~\ref{fig_compare_empiric}b). 
We observe a significant decline of this number, which rules out the possibility that the empiric case corresponds to the domination of local networks. Because the past can be considered a single realization of a stochastic process~\cite{watts:obvious}, the empiric case can still be within the coinflip region of our model where --by chance-- the international network was more successful. Hence, we will perform the following comparison only for realizations of our model in which local networks do not dominate.

The intrinsic timescale of the model is arbitrary and hence has to be mapped to real time. The optimal mapping is given such that it produces the best agreement with the empirical data. We quantify the agreement between model results and empirical data using the sum of the squared distances between the data points and model results. In particular, we use the $\chi^2$ statistic defined as:
\begin{equation}
 \chi^2 =   \frac{1}{\sigma_N^2}\sum_i \left[N_i - N_i^{\text{model}} \right]^2 \eqcomma
 \label{eqn_capital_gamma}
\end{equation}
where $N_i$ denotes the number of countries where the local network is more popular and $N_i^{\text{model}}$ is the corresponding result from the model. The index $i$ denotes the individual datapoints and $\sigma_N^2$ is the estimated variance of the data (Supplementary Materials).
The real time, $t_R$, is a linear function of the model time $t_M$ given by $t_R = a_1 + a_2 t_M$, where $a_1$ is the starting year and $a_2$ represents the time stretch: how many years of real time correspond one model time step.
For a given set of parameters, $\sigma$, $\alpha$, and $\Delta t$, the optimal values for $a_1$ and $a_2$
are those that minimize $\chi^2$, as shown in Fig.~\ref{fig_compare_empiric}a. 

We can also use the $\chi^2$ statistic to estimate the parameters $\alpha$, $\sigma$, and $\Delta t$ which best reproduce the empirical observations. In Fig.~\ref{fig_compare_empiric}c, we plot the values of $\chi^2$ as a function of $\alpha$, $\sigma$, and $\Delta t$, where --at each point-- we applied the respective best time mapping, as described above. These results are averaged over several realizations of the model; however, in the coinflip region we exclude realizations where the local networks dominate, to mimic the empirical case.
Interestingly, the overall best fit is achieved for $\alpha = 2$ at $\sigma=1.25$ and $\Delta t = 3$, which lies in the coinflip region
(the optimal value of $\chi^2$ is statistically consistent with the model, given the number of degrees of freedom in the data) with a probability for domination of the international network of $70\%$.
This scenario corresponds to the time mapping $a_1 = 2006$ and $a_2 = 0.6$, meaning the system started at the beginning of $2006$; while the launch time delay of $\Delta t = 3$ in the model translates to $1.8$ years in real time. 
In Fig.~\ref{fig_compare_empiric}b, we show the evolution of the number of countries where local networks are more popular for the optimal fit from the model.

\section{Discussion}

Understanding the complex dynamics of the digital world constitutes an important challenge for interdisciplinary science. 
To meet this challenge, here we describe the the worldwide web 
as a complex, digital ecosystem in which interacting networks play the role of species in competition for survival. In particular, we study the competition between local networks operating in single countries and an international network that operates in all countries.
Therefore, a proper description of this system must necessarily involve the network of worldwide social interactions between different countries.

We show that the effect of inter-country social ties can be mapped to the increased fitness of the international network by means of an effective activity.
Interestingly, there is a critical global coupling strength below which networks can coexist.
However, above that threshold, only domination is possible: in general, local networks become extinct with a high probability.
Yet, we find that if local networks are launched earlier they can persist and dominate the international network, which happens only if local networks have accumulated a sufficiently large active userbase when the global launch of the international network takes place. 
The accumulation of a sufficient base depends on the parameters; and for certain parameters on chance. For these parameters the final state of the system --whether local networks dominate or become extinct-- can be completely unpredictable, as it varies randomly between different realizations of the model.
 
Quite remarkably, a thorough comparison of our model with empirical data from the recent takeover of Facebook indicates that the most probable launch date of Facebook was at the beginning of $2006$ and its global launch was in late $2007$. Facebook was in fact started in $2004$, but opened to the public in $2006$; in good agreement with the estimate from our model. Moreover, according to Google trend data (see Supplementary Materials), $2007$ was the year when the global search volume for Facebook started to increase rapidly. 
Last but not least, our best estimation of the model parameters corresponds to the ``coinflip'' region, which means that 
the observed takeover of Facebook only had a probability of around $70\%$.
With a $30\%$ probability, we would have been living in a world where each country had its own successful local network and a network like Facebook would not exist~\cite{watts:obvious}.

Our findings suggest interesting future lines of research. On the one hand,
even without adjusting the parameters on a country-by-country level, our model reproduces the main features empirically observed in the takeover of Facebook and the extinction of local networks in most countries for a certain parameter region. 
It remains an interesting task for future research to further increase the precision of the model. This could be done by improving the proxy for the similarity between countries or by adjusting parameters on a country-by-country basis. 
On the other hand,  
the model could be extended to account for several international networks and to study their global competition. For a second international network to overcome the first, a certain minimal difference of fitness is needed; which could be the result of different properties of the networks, such as features or functionalities. 
Finally, random fluctuations of fitness could be incorporated to describe Darwinian selection in the digital ecosystem.

\section{Methods and Material}

\subsection{S1 model}
\label{app_sec_s1}

We use the $\mathbb{S}1$ model~\cite{Serrano2008,Boguna2008,Boguna2003} to generate the synthetic networks for the underlying societies in each country. The model allows us to specify the degree distribution and the level of clustering. The model is based on a circle as a hidden metric space and works as follows:
\begin{enumerate}
 \item All nodes are placed on the circle with a randomly assigned variable, $\theta$, which represents the polar coordinate. $\theta$ is uniformly distributed in $[0,2\pi)$. To keep the average node density on the circle constant, its radius grows linearly with the number of nodes, to satisfy $N = 2 \pi R$. 
 \item We assign each node a second hidden variable, $\kappa$, which represents its expected degree. $\kappa$ is drawn from an arbitrary distribution $\rho(\kappa)$. 
 \item A pair of nodes is connected with a probability, $r$, that depends on their hidden variables $(\theta,\kappa)$ and $(\theta',\kappa')$:
 \begin{equation}
 r(\theta,\kappa;\theta',\kappa') = \left(1+\frac{d(\theta,\theta')}{\mu \kappa \kappa'}\right)^{-\alpha} \eqcomma
\end{equation}
with $\mu = \frac{\alpha-1}{2 \km}$. Here, $d(\theta,\theta')$ denotes the geodesic distance between the two nodes on the circle and $\km$ the mean degree. Then, the expected degree, $\bar{k}(\kappa)$, of a node with hidden variable $\kappa$ can be shown to be proportional to $\kappa$~\cite{Boguna2003}. As a consequence, the degree distribution, $p(k)$, of the network follows the shape of the distribution $\rho(\kappa)$. 
\end{enumerate}
\begin{figure}[b]
\centering
 \includegraphics[width=1\linewidth]{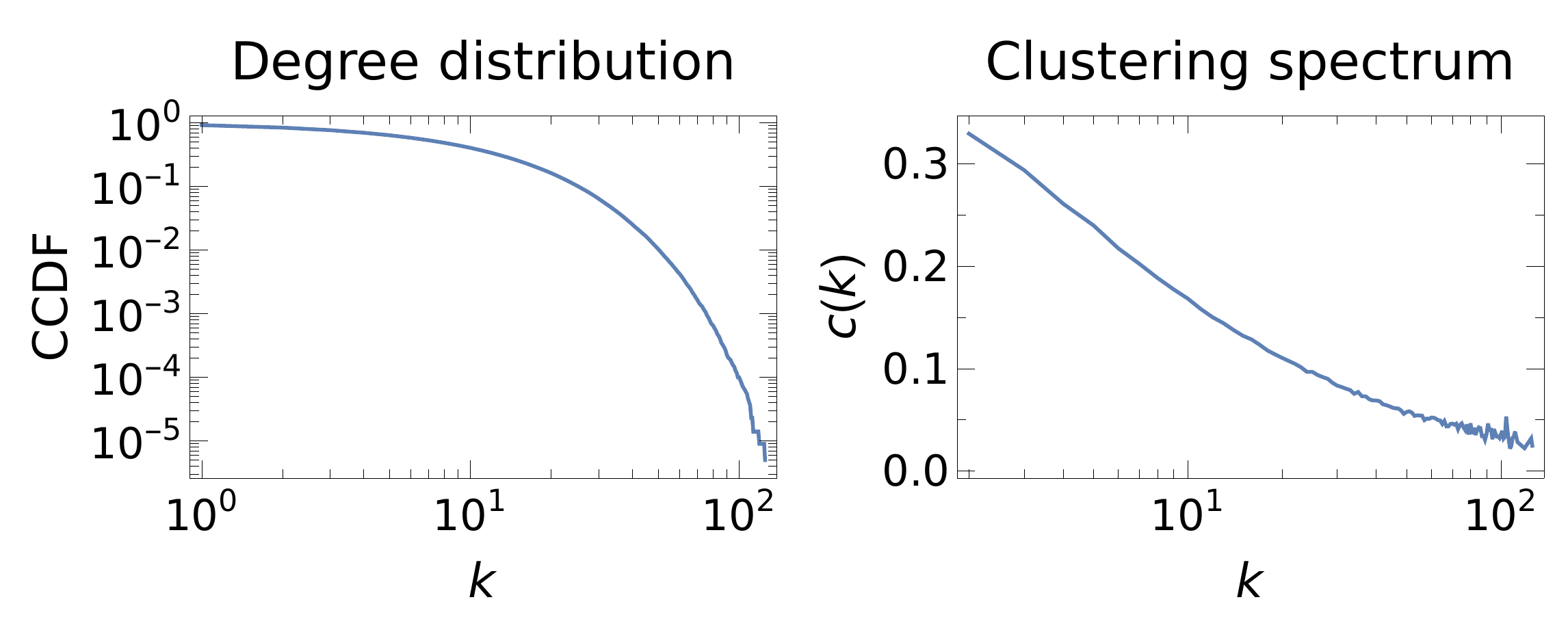}
 \caption{Degree distribution and clustering spectrum for networks generated for the example of the US ($\approx 230.000$ nodes). 
 \label{fig_s1_properties}}
\end{figure}
Here, we use an exponential distribution $\rho_{\xi}(\kappa) = \xi e^{-\xi \kappa}$ with $\xi = 10$. We set the parameters $\alpha = 1.5$ and $\mu = 0.02$. After generating the networks, we remove nodes with zero degree. Fig.~\ref{fig_s1_properties} shows the degree distribution and the clustering spectrum for the synthetic network created for the US.

\subsection{Air travel data}
\label{sec_app_data}
Air travel data aggregated on a country basis were taken from 
http://visualising.org/datasets/global-flights-network
(date of access July 2014). The original data can be accessed at http://openflights.org/data.html. The network on a country basis incorporates 230 nodes and 4600 weighted edges which correspond to the number of routes between countries, i.e. the number of total flights. 
The dataset contains around 60.000 of such flights.
We extract the subnetwork (see Fig.~\ref{fig_panel1}c) by constraining to the countries listed in Tab.~\ref{tab_countries}.

\subsection{Supplementary Material}
Supplementary Material is available at: http://www.nature.com/article-assets/npg/srep/2016/160427/srep25116/extref/srep25116-s3.pdf

\begin{table*}[t]
\begin{ruledtabular}
 \begin{tabular}{llllllllll}
 China & 253. & UnitedStates & 231. & Japan & 90.9 & India & 81. & Brazil & 64.9 \\
 Germany & 62. & UnitedKingdom & 48.8 & Russia & 45.2 & France & 42.9 & SouthKorea & 37.5 \\
 Indonesia & 30. & Spain & 25.2 & Canada & 25.1 & Italy & 25. & Turkey & 24.5 \\
 Mexico & 23.3 & Iran & 23. & Vietnam & 20.8 & Poland & 18.7 & Pakistan & 18.5 \\
 Colombia & 17.1 & Malaysia & 16.9 & Thailand & 16.1 & Australia & 15.2 & Taiwan & 15.1 \\
 Netherlands & 14.3 & Egypt & 11.4 & Argentina & 11.2 & Nigeria & 11. & Ukraine & 10.4 \\
 Morocco & 10.3 & Sweden & 8.1 & SaudiArabia & 7.7 & Belgium & 7.3 & Venezuela & 7.2 \\
 Peru & 7.1 & Romania & 6.1 & CzechRepublic & 6. & Austria & 5.9 & Hungary & 5.9 \\
 Switzerland & 5.7 & Philippines & 5.6 & Chile & 5.5 & Denmark & 4.6 & Portugal & 4.5 \\
 Finland & 4.4 & Greece & 4.3 & Sudan & 4.2 & SouthAfrica & 4.2 & HongKong & 4.1 \\
 Algeria & 4.1 & Norway & 3.9 & Slovakia & 3.6 & Syria & 3.6 & Singapore & 3.4 \\
 Kenya & 3.4 & Belarus & 3.1 & NewZealand & 3. & Serbia & 2.9 & UnitedArabEmirates & 2.9 \\
 Ireland & 2.8 & Tunisia & 2.8 & Bulgaria & 2.6 & Uganda & 2.5 & Uzbekistan & 2.5 \\
 Kazakhstan & 2.3 & Lebanon & 2.2 & DominicanRepublic & 2.1 & Israel & 2.1 & Guatemala & 2. \\
 Croatia & 1.9 & Lithuania & 1.8 & Jamaica & 1.5 & Jordan & 1.5 & Azerbaijan & 1.5 \\
 CostaRica & 1.5 & Cuba & 1.4 & Zimbabwe & 1.4 & Uruguay & 1.3 & Ecuador & 1.3 \\
\end{tabular}
\end{ruledtabular}
\caption{List of countries and estimated number of Internet users ($\times 10^6$) according to Wolfram Alpha database.\label{tab_countries}}
\end{table*}

\begin{acknowledgments}
This work was supported by: the European Commission within the Marie Curie ITN ``iSocial'' grant no.\ PITN-GA-2012-316808; a James S. McDonnell Foundation Scholar Award in Complex Systems; the ICREA Academia prize, funded by the {\it Generalitat de Catalunya}; the MINECO projects nos.\ FIS2010-21781-C02-02 and FIS2013-47282-C2-1-P;  and the {\it Generalitat de Catalunya} grant no.\ 2014SGR608. Furthermore, M.~B. acknowledges support from the European Commission LASAGNE project no.\ 318132 (STREP).
\end{acknowledgments}

% 
% \bibliography{mybib}
% \bibliographystyle{unsrt}

\end{document}